# Design and performance of an ultrahigh vacuum spectroscopic-imaging scanning tunneling microscope with a hybrid vibration isolation system

Pei-Fang Chung,[1,a] Balaji Venkatesan,[1,2,3,a] Chih-Chuan Su,[1,a] Jen-Te Chang,[1] Hsu-Kai Cheng,[1] Che-An Liu,[1] Henry Yu,[1] Chia-Seng Chang[1,2], Syu-You Guan[1] and Tien-Ming Chuang[1,b,*]

[1] *Institute of Physics, Academia Sinica, Taipei 11529, Taiwan*

[2] *Department of Physics, National Taiwan University, Taipei 10617, Taiwan*

[3] *Nano Science and Technology Program, Taiwan International Graduate Program, Academia Sinica and National Taiwan University, Taipei 11529, Taiwan*

(*Electronic mail: chuangtm@gate.sinica.edu.tw)

A spectroscopic imaging-scanning tunneling microscope (SI-STM) allows the atomic scale visualization of surface electronic and magnetic structure of novel quantum materials with high energy resolution. To achieve the optimal performance, low vibration facility is required. Here, we describe the design and the performance of an ultrahigh vacuum STM system supported by a hybrid vibration isolation system that consists of a pneumatic passive and a piezoelectric active vibration isolation stages. The STM system is equipped with a 1K pot cryogenic insert and a 9 Tesla superconducting magnet, capable of continuous SI-STM measurements for 7 days. A field ion microscopy system is installed for *in situ* STM tip treatment. We present the detailed vibrational noise analysis of the hybrid vibration isolation system and demonstrate the performance of our STM system by taking high resolution spectroscopic maps and topographic images on several quantum materials. Our results establish a new strategy to achieve an effective vibration isolation system for high-resolution STM and other scanning probe microscopy to investigate the nanoscale quantum phenomena.

## I. INTRODUCTION

Spectroscopic-imaging (SI) scanning tunneling microscopy (STM) has become a powerful tool for directly visualizing novel quantum phenomena at the surface with high spatial and energy resolution[1,2]. The SI-STM measurement yields a differential conductance map, $dI/dV(\bm{r}, E)$ that reveals the spatial distribution of local density of state (LDOS) of a material as a function of energy, e.g. superconducting gap[3] or vortex maps[4,5]. Fourier transform of the $dI/dV(\bm{r}, E)$ map can extract the periodic modulation of LDOS in real-space from quasiparticle scattering interference (QPI) and allows the determination of the momentum-space electronic structure of a material[6-9]. Furthermore, the spin[10,11], orbital[12] and phase[13,14]-sensitive QPI process provides further insight to the properties of novel quantum materials, such as unconventional superconductors and topological materials. The changes in the electronic structure driven by the external magnetic field[15] or strain[16,17] can also be imaged. The unique capability to obtain energy resolved information on the wavefunction and energy dispersion of quasiparticles in both real and momentum space makes SI-STM a powerful technique for material research. Recent developments also extend the STM capability to detect electron spin resonance[18,19] and shot noise[20-23] at atomic scale. Thus, SI-STM is indispensable for investigating and developing new advanced materials.

Signal-to-noise ratio and stability are crucial to optimize the energy resolution of SI-STM. The tunneling current is prone to vibration noise during the SI measurement since the feedback loop is open at each pixel of a spectroscopic map, which can take days to finish. To achieve such conditions, it is necessary to shield the STM from the detrimental floor vibration and acoustic noise during the SI measurement. Additionally, it is also desirable to retain the register to the same field of view (FOV) as we tune the experimental parameters such as temperature or magnetic field. In the past decades, ultralow vibration facilities for housing STMs and other sensitive instrument have been constructed on the basis of a stacked pneumatic-pneumatic vibration isolation system[24-30]. In such a design, large pneumatic passive vibration isolators support a heavy concrete plinth weighted tens of tons to decouple from the building, on top of which a smaller pneumatic passive vibration isolation stage supporting the instrument is installed. Typical pneumatic spring-mass system can effectively reduce vibration noise above the resonant frequency of the whole system in the range of 1-10 Hz. Thus, it is important to design the mass ratio between the two vibration isolation stages and to decouple the resonance frequencies of two pneumatic isolators to avoid the vibration amplification and the performance loss[31]. In addition, the center of mass of both stages in a stacked system need to be close to the supporting plane of isolators to ensure the tilt stability. Moreover, it is very challenging to modify an existing building to accommodate such a massive system. It is also expensive and time-consuming to construct such a laboratory before the instrument can be installed and optimized. Effective vibration isolation strategy to mitigate these issues is highly desired for SI-STM, advanced metrology and nanofabrication applications.

Previously, Iwaya *et al.* has presented a detailed vibration analysis of a STM system with a hybrid active-passive vibration isolation system, providing a solution to overcome these issues[32]. Here, we present the design and the construction of a home-build ultrahigh vacuum (UHV) STM system supported by a hybrid vibration isolation system, in which a passive pneumatic passive stage stacks on top of a piezoelectric active stage, and housed inside an acoustic shielded room. A UHV field ion microscopy (FIM) is used to treat and

---

a) These authors contributed equally to this work.
b) Author to whom correspondence should be addressed

characterize the STM tips *in situ* before measurements. The STM head is mounted on a 1K pot based cryogenic insert, which can be operated in a temperature range between T = 1.5 K and 60 K in a magnetic field of up to 9 Tesla normal to the sample surface for up to 7 days. We perform systematic vibrational noise measurements and show that this hybrid vibration isolation system is effective to suppress the vibration noise below ~ 100 Hz. Finally, we demonstrate the performance of our STM by taking topographic and SI measurements on several quantum materials.

## II. SYSTEM OVERVIEW

### A. Vibration isolation system

Figure 1 shows the overview of the entire homebuilt UHV STM system located in the basement of the Institute of Physics, Academia Sinica. The laboratory floor, which is built on a raft foundation, connects directly to neighboring laboratories and large service facilities including water chiller, ventilation and air conditioning system. The laboratory is also located only 20 meters away from the traffic and bus routes. Because of the technical constraints, we are unable to construct a dedicated low vibration laboratory decoupled from the rest of the building. Therefore, we design a hybrid vibration isolation system to be installed directly on the floor. The system consists of two stages of vibration isolation. The first stage is a stiff SS304 platform (185 cm × 198 cm × 10.16 cm, custom-made STACIS Floor Platform, TMC-Ametek) supported by four piezoelectric active isolators (STACIS III, TMC-Ametek). A hole with a diameter of 75 cm is also made on the platform so the dewar can be lowered into a pit on the floor during the maintenance. Above the first stage, we construct a second vibration isolation stage, which is comprised of three non-magnetic pneumatic passive isolators (natural frequency ~ 2 Hz, custom-made Gimbal Piston, TMC-Ametek) bolted onto three lead shot - filled pillars (weighted ~ 500 kg each). A triangular table made of welded aluminum beams and filled with ~ 900 kg of lead floats above these passive isolators. The entire STM system, consisting of a super-insulated liquid helium dewar (60 liters capacity, Kadel Engineering Co.) with a 9 Tesla superconducting magnet (American Magnetics, Inc.), a homemade 1 K pot based cryogenic insert with the STM head (Fig. 1) and our UHV chamber (Fig. 2a), is then mounted on the triangular table. Our STM system is housed inside an acoustic shield room of noise isolation class 45 to reduce the acoustic noise from the environment. All roughing pumps and electronics, except for the current preamp, are placed outside the acoustic shield room to minimize their perturbation during the measurement. Additionally, a back pressure regulator (Model 710-DA, ControlAir) is installed at the exhaust of the dewar to maintain the vapor pressure of the liquid bath at ~ 0.5 psi above the atmosphere pressure, which greatly stabilizes the base temperature and also prevents the noise from the downstream of the helium recovery line.

### B. UHV chamber and field ion microscopy system

A UHV chamber is designed and mounted on the triangular table for the tip/sample transfer, the tip preparation and the tip characterization as shown in Fig. 2(a). It is comprised of a load-lock chamber and a main chamber with a field ion microscopy (FIM). The load-lock chamber is fitted with a quick access door and a turbomolecular pump to allow the sample and the tip to be transferred quickly in and out of the main chamber and then the STM head at the low temperature. We design a shuttle that can simultaneously carry two sample holders and six tip holders, enabling the efficient sample and tip transfer (Fig. 2b).

The UHV main chamber is equipped a homemade FIM system to prepare a clean STM tip *in situ* by field evaporation and to analyze it by imaging. The main chamber is equipped with a titanium sublimation combination ion pump (Vaclon plus 150, Agilent) for maintaining the UHV (typically $< 4 \times 10^{-10}$ mbar), an ion gauge for vacuum measurement and a residual gas analyzer (RGA100, Stanford Research Systems) for leak detection. Several manipulators are installed for sample/tip transfer. Extra ports are reserved for adding more sample preparation apparatus in the future.

Figure 2 (b) depicts the workflow for the tip treatment and the tip/sample transfer. First, the shuttle carrying two sample holders and six tip holders is moved from the load-lock chamber to the main chamber by using the X-manipulator and then picked up by the Y-manipulator. Second, the STM tip is loaded into the ceramic FIM stage on the Y-manipulator from the shuttle by using a tip grabber, which is installed on the Z-manipulator. Third, the Y-manipulator rotates 90° to accurately orient the STM tip towards the microchannel plate (MCP, Model F1552, Hamamatsu) with a pre-determined distance for tip preparation and FIM imaging. Last, the clean tip is picked up by the tip grabber on a 1.6-meter-long vertical manipulator and is transferred into the STM head at the low temperature. Similarly, the sample is transferred to the STM head by the sample grabber on the vertical manipulator. Tips and samples are pre-cooled at the 4 K plate and single crystals are cleaved at T < 20K by using a rotary cleaver, which also serves as radiation shield and the sample storage, before being inserted to the STM head for measurements.

### C. Cryogenic insert

A homemade cryogenic insert is designed to house the STM and the related operations. The insert is rigidly connected with the UHV chamber on the triangular table but is separated by a gate valve, which is closed after cooled down because the cryogenic pumping from the liquid helium bath is more effective than an ion pump. The bottom of the insert is coupled to the room temperature by a center tube and six tubes with a hexagonal pattern with respect to the center one, which are all made of electro-polished 0.010" thin walled SS316 tubes (MicroGroup Inc. and Eagle Stainless Tube & Fabrication, Inc.) to minimize the heat load. The center tube is used for transferring the tip/sample and pumping the insert. Among the six outer tubes, one is used for pumping the 1 K pot, one is used for mechanical heat switch that connects the 4 K plate and the 1 K stage during the initial cool down from room temperature, one is used for the manipulator that controls the cleaver, the cryogenic sample storage and the radiation shield and three are used for wiring.

To further reduce the heat transfer, the wiring for temperature control and piezoelectric coarse approach use twisted phosphor bronze wires (Lakeshore) and the sensitive control/signal lines for the STM operation use CuNi cladded superconducting NbTi coax cables with SS316L braided shield (Cooner Wires). All wires are carefully thermally anchored at the 4K and the 1K plates made of gold-plated oxygen free copper. Radiation baffle plates are also fitted at the 4 K plate inside all six tubes. To reduce the radio frequency noise, all wires are connected through π filters except the tunneling current. We use the π filters (P/N 4209-053, Tusonix Inc.) for the bias and the high voltage piezo lines and the π filters (P/N 51-712-065, API Technologies.) for other low voltage lines. As a result of careful design, we achieve a static helium boil-off rate of ~ 6.8 liters / day at T = 4.2 K (compared with 4.8 liters / day without the insert), allowing continuous SI-STM measurements for ~ 7 days.

The design of the 1 K cooling stage consists of a 1 K pot with the volume of 18 cc and a flow impedance made of a 2-meter-long CuNi capillary with inner diameter of 0.1 mm. The capillary is wrapped around a copper post and thermally anchored at the 1 K plate before the incoming liquid helium enters the pot to reduce the noise. When the STM experiment needs to be performed at T > 20 K, the helium vapor with the 1 K line significantly increases the heat transfer between the STM and the 4K plate, which is directly immersed in the liquid helium bath. In this case, the flow impedance is removed before cool-down and the 1 K line is evacuated with a mechanical pump to eliminate this heat transfer. This allows us to achieve the helium boil-off rate of ~ 7.4 liters per day when the 1 K stage and the STM head are at T ~ 40 K. In comparison, the boil-off rate increases to 14 liters per day when the flow impedance is not removed.

### D. STM Head

Our STM is based on the Pan-style walker design due to its excellent rigidity and stability so it can be directly mounted on the bottom of the 1 K stage without internal damping as shown in Fig 1(b). The detail construction of our STM head is illustrated in Fig. 3. The STM body (outer diameter of 35 mm) and most of insulating components are made of machinable ceramic Shapal due to their similar thermal expansion with piezoelectric materials, high mechanical strength and relatively good thermal conductivity. Six piezostacks, each of which is made of four shear piezo plates (PIC 255, Physik Instrumente) and a polished alumina plate, drive a polished sapphire prism (Sarton Works Co., Ltd.) that supports the tip holder and the piezoelectric scanner (PZT - 8, EBL Products, Inc.). The scanner has a scan range of $650 \times 650$ nm$^2$ and $z \sim \pm 110$ nm at T ~ 4.2 K with applied voltage of $\pm 220$ V.

The clamping force for the walker is provided by a BeCu spring plate with a torque of ~ 2 cN – m[33]. The walker is driven in a stick-slip mode by using a Nanonis PMD4 controller and achieves a step size of ~ 40 nm at T ~ 4.2 K when a bipolar sawtooth-like waveform of $\pm 110$ V is applied. A capacitive position sensor is glued to the bottom of the scanner so that the position of the scanner and the STM tip can be measured by a capacitance bridge (Model AH2550, Andeen-Hagerling Inc.). The tip – sample distance can also be detected by measuring the capacitance between the tip and the sample holder with a lock-in amplifier. The tip holder, essentially a M3 screw, is inserted into a matching receptacle glued on the top of piezo scanner tube by using the vertical UHV transfer rod. The sample stage is made of titanium with a BeCu spring, which also serves as the electrode for the sample bias in addition to securing the sample holder. Before replacing the STM tip, the sample needs to be removed from the sample stage and moved to the sample storage at the 4 K plate first. The tunneling current is amplified with a gain of $10^9$ by a wide bandwidth pre-amplifier (SA-606F2, nF Corp.) and is acquired by using a Nanonis RC4/SC4 controller.

## III. PERFORMANCE OF THE INSTRUMENT

### A. Vibration and current noise measurements

To evaluate the performance of our hybrid vibration isolation system, we activate the piezoelectric active isolators and pneumatic passive isolators separately and then measure the vibration noise by using a geophone (Model GS-1, Geospace Technologies.) at the triangular table. Figure 4 (a) - (c) and supplemental figure show the comparison between the vibration noise spectrum and the transfer function along the x, y, and z directions, respectively. When both isolators are off, the vibration noise spectrums in three directions are similar, except higher noise between 5 Hz and 10 Hz in x- and y-direction can be observed.

When only the piezoelectric active isolators are turned on, the vibration level in all three directions are greatly reduced below ~ 40 Hz, but are much less effective above ~ 50 Hz and ~ 60 Hz for horizontal and z-direction, respectively. When only the pneumatic passive isolators are turned on, the vibration levels in all directions are enhanced inevitably below 3 Hz due to the natural frequency of the passive isolators. The vibration levels are reduced similarly below 40 Hz, except around 11Hz and 20 Hz, which are likely the structural resonance of the triangular table and frame. In contrast, they remain effective to lower the vibration above 60 Hz, especially in the z-directions.

As both isolators are turned on, those higher vibrational peaks between 3 Hz and 30 Hz when active and passive isolators are running separately are suppressed in all directions. The vibration levels above 60 Hz are also reduced to the similar level as the case of only the passive isolators turned on. We find both isolators are less effective between 40 Hz and 60 Hz, where the highest noise levels are ~ $5 \times 10^{-8}$ m/sec/√Hz. We deduce the combination of piezoelectric active isolators and pneumatic passive isolators gives complementary vibration isolation performance from the natural frequency of pneumatic passive isolators up to ~100 Hz, except between 40 Hz and 60 Hz. We note the large vibration from the surrounding service facilities occurs between ~ 25 and ~ 70 Hz, which can be seen from the comparison of the geophone measurements between the regular operation and the annual maintenance when all heavy machinery nearby is shut down (the inset of Fig. 4 (d)).

We further test the noise level of our STM by taking the tunneling current spectrum on a 15-nm-thick superconducting δ-NbN film grown on MgO (Tc ~ 16 K) with a W tip at T ~ 4.2K. The feedback is turned off after the tunneling current is stabilized at a current setpoint of 100 pA with a bias of 50 mV. No obvious noise peaks are visible from DC to 1 kHz except a relatively small noise at 60 Hz due to electrical power line, as shown in Fig. 4(d). The lack of correlation between the tunneling current and the vibration noise spectrum is attributed to our rigid STM construction.

### B. SI-STM performance

Polycrystalline tungsten tips are cleaned and characterized at the FIM stage before used for STM measurements. Helium gas of ~ $1 \times 10^{-4}$ mbar is ionized near the tip apex with an applied high voltage of maximum 10 kV to the tip and accelerated toward the MCP for imaging on a phosphor screen. The tip can be sufficiently cleaned during the imaging process without prior annealing, which is evident from the clear FIM image shown in Fig. 5 (a). The tip sharpness can also be estimated by the applied voltage to the tip[34].

To validate the performance of our STM system, we first measure the type-II Weyl semimetal $WTe_2$ single crystals. The sample is pre-cooled at the 4 K plate and then cleaved at T < 20 K. Then, the sample is immediately transferred into the STM head at T ~ 4.2 K. The high resolution topographic image of $WTe_2$ taken in a $10 \times 10$ nm$^2$ FOV is shown in Fig. 5(b), which reveals its characteristic chain structure of surface Te atoms due to the orthorhombic distortion. A topographic image taken at a $60 \times 60$ nm$^2$ FOV in Fig. 5(c) shows large flat cleaved surface with two different impurities. Each of which produces a distinct QPI pattern in real space during the differential conductance mapping measurement, $dI/dV(\mathbf{r}, E)$, which is evident in Fig. 5(d). Fourier-transform of $dI/dV(\mathbf{r}, E)$ maps yield dI/dV($\mathbf{q}$, E) maps (unprocessed in Fig. 5(e)), which can be compared with the band structure and is consistent with previous QPI imaging results on $WTe_2$[35-37]. We also measure the superconducting gap of the δ-NbN film at T ~ 1.5 K. The high resolution spectrum reveals a superconducting gap of ~ 4.44 meV and can be fitted well by using a Dynes function as shown in Fig. 5(f).

## IV. SUMMARY

In summary, we have designed and constructed a homebuilt UHV STM system, supported by a hybrid vibration isolation system with stacked piezoelectric active and pneumatic passive isolators. We have performed the vibration noise measurement and the results have demonstrated that this design can suppress the vibration in all directions up to 100 Hz. The high rigidity of our STM head can further mitigate the residual low frequency vibration noise as evident from the current noise spectrum. Our data on the type - II Weyl semimetal $WTe_2$ and the superconducting δ-NbN film demonstrate the performance of our system, capable of spectroscopic imaging investigation of quantum materials at the atomic scale. Our results provide an efficient vibration isolation method tailored for high resolution scanning probe microscopy and are also of great importance to advanced nanofabrication and quantum metrology applications.

## ACKNOWLEDGMENT

We are grateful for the helpful discussion with Tetsuo Hanaguri, J. C. Séamus Davis, Shuheng Pan, Andreas W. Rost, Milan P. Allan, Peter Wahl, Mohammad Hamidian, Chih-Kang Ken Shih, Jhinhwan Lee and Ing-Shouh Hwang. We thank the financial support from the National Science and Technology Council, Taiwan (NSTC 112-2112-M-001-046-MY3, NSTC 111-2112-M-001-024-MY2) and Academia Sinica, Taiwan (AS-iMATE-111-12). We acknowledge the machine shop staff at the Institute of Physics, Academia Sinica, whose craftsmanship was instrumental in realizing our design. We also thank TMC-Ametek and Scientech Corp. for their technical assistance, Raman Sankar for supplying high quality $WTe_2$ single crystals and Yu-Jung Yuri Lu for providing us NbN thin film samples. T.M.C acknowledges the generous support from Golden-Jade Fellowship of Kenda Foundation, Taiwan.

## AUTHOR DECLARATIONS

### Conflict of Interest

The authors have no conflicts to disclose.

### Author Contributions

**Pei-Fang Chung**: Conceptualization (equal); Data curation (lead); Investigation (equal); Writing - original draft (equal); Writing – review & editing (equal). **Balaji Venkatesan**: Conceptualization (equal); Data curation (lead); Investigation (equal); Writing – review & editing (equal). **Chih-Chuan Su**: Conceptualization (equal); Data curation (lead); Investigation (equal); Writing – review & editing (equal). **Jen-Te Chang**: Data curation (equal); Investigation (equal). **Hsu-Kai Cheng**: Data curation (equal); Investigation (equal). **Che-An Liu**: Data curation (equal); Investigation (equal). **Henry Yu**: Data curation (equal); Investigation (equal). **Chia-Seng Chang**: Conceptualization (equal); Writing – review & editing (equal); Supervision (supporting). **Syu-You Guan**: Conceptualization (equal); Data curation (equal); Investigation (equal); Writing – review & editing (equal). **Tien-Ming Chuang**: Conceptualization (lead); Investigation (equal); Supervision (lead); Writing - original draft (equal); Writing – review & editing (equal).

## DATA AVAILABILITY

The data that support the findings of this study are available from the corresponding authors upon reasonable request.

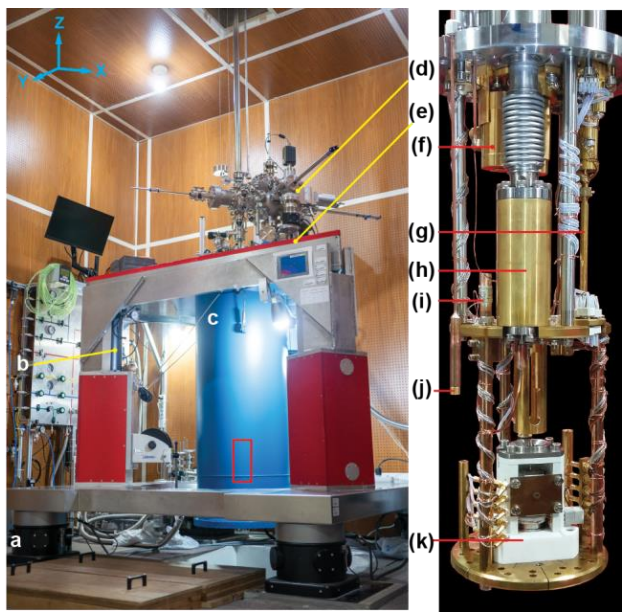

**Figure 1**. Overview of the entire homebuilt UHV STM system inside the acoustic shield room. It consists of two stages of vibration isolation. The first stage of four piezoelectric active isolators (a) supports the second stage of three non-magnetic pneumatic passive isolators (b). A 9-Tesla superconducting magnet dewar (c), a UHV chamber (d) and the cryogenic insert with the STM head (right) are mounted on the triangle table (e) filled with ~900kg of lead and supported by the pneumatic passive isolators. The detail of the cryogenic insert (red rectangle) is shown in the right figure: the cleaver, cryogenic sample storage and the radiation shield (f), a mechanical heat switch (g) between the 4 K plate and the 1 K plate, the 1 K pot (h), the flow impendence (i), the intake of impendence (j), and the STM head (k).

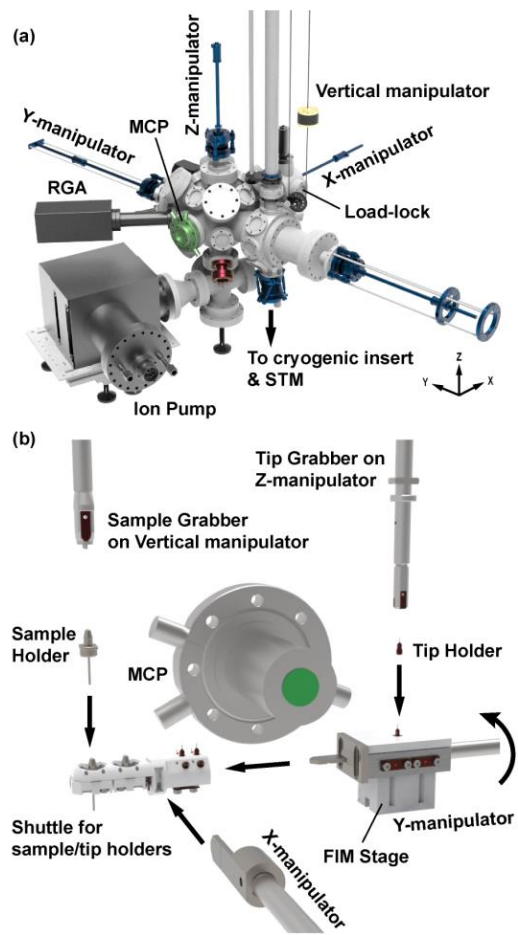

**Figure 2**. (a) Schematic of the UHV main chamber, which consists of a load-lock chamber, a residual gas analyzer (RGA), an ion pump and a field ion microscopy system. (b) Tip preparation/characterization and tip/sample transfer. The shuttle with two sample holders and six tip holders is inserted from the load-lock chamber by X-manipulator and picked up by Y-manipulator. The tip holder is transferred to the FIM stage by the Z-manipulator and then rotated to point the tip toward the MCP for field ion microscopy experiment. A sample grabber (a tip grabber) installed on the vertical manipulator is used to transfer the sample (the tip holder) to the STM head.

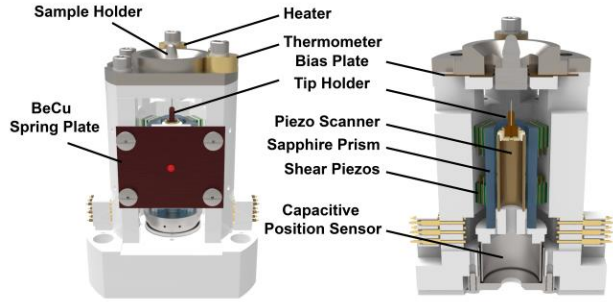

**Figure 3.** (Left) The 3D rendering of the STM head. The outer diameter of STM is 35mm with a height of 70mm. (Right) The cross-section view of the STM head shows the interior construction of the walker for coarse approach, the piezo scanner assembly, and the sample stage.

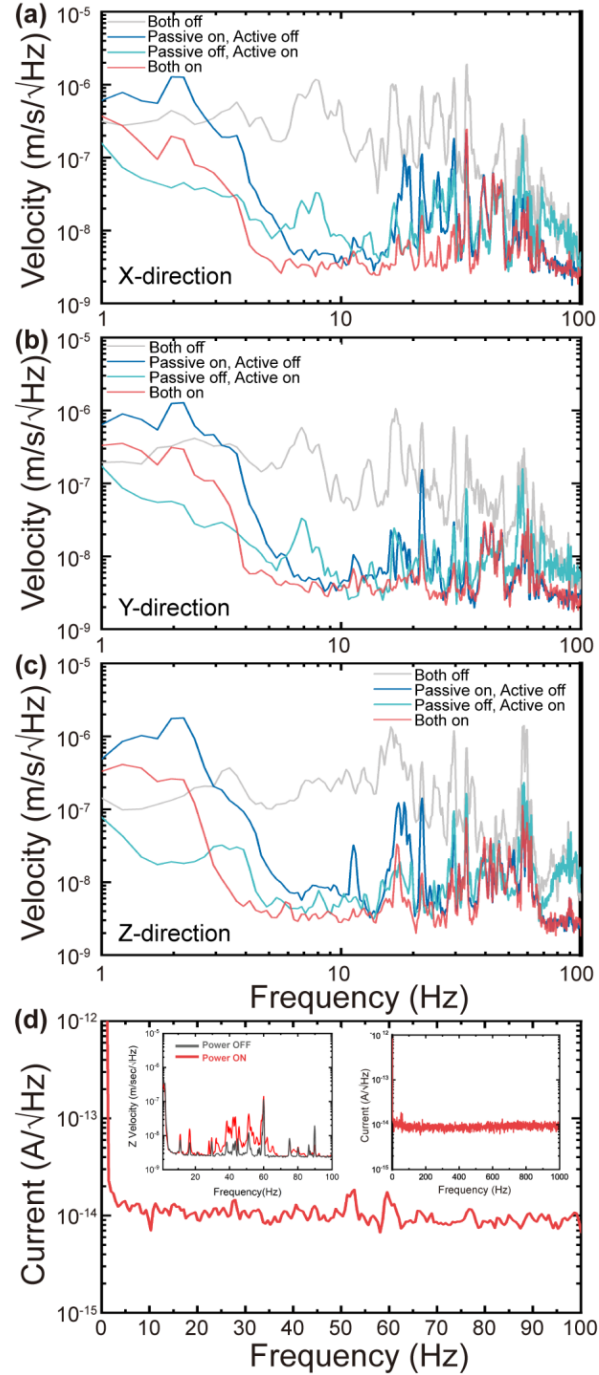

**Figure 4.** The vibration noise measurements of our hybrid vibration isolation system taken at the top of the triangle table along (a) X-direction, (b) Y-direction, and (c) Z-direction, respectively. (d) The tunneling current noise spectrum between 0 to 100 Hz taken on a NbN thin film at T~ 4.2K with open feedback loop (sample bias voltage, V = 50 mV and current setpoint, I = 100pA). The right inset shows the current noise spectrum between 0 and 1 kHz. The left inset shows the vibration comparison along the z-direction when all surrounding facilities stop running (power OFF, black curve) and during the regular operation (power ON, red curve). The data represents the averaged results of an 8 hour long continuous geophone measurement.

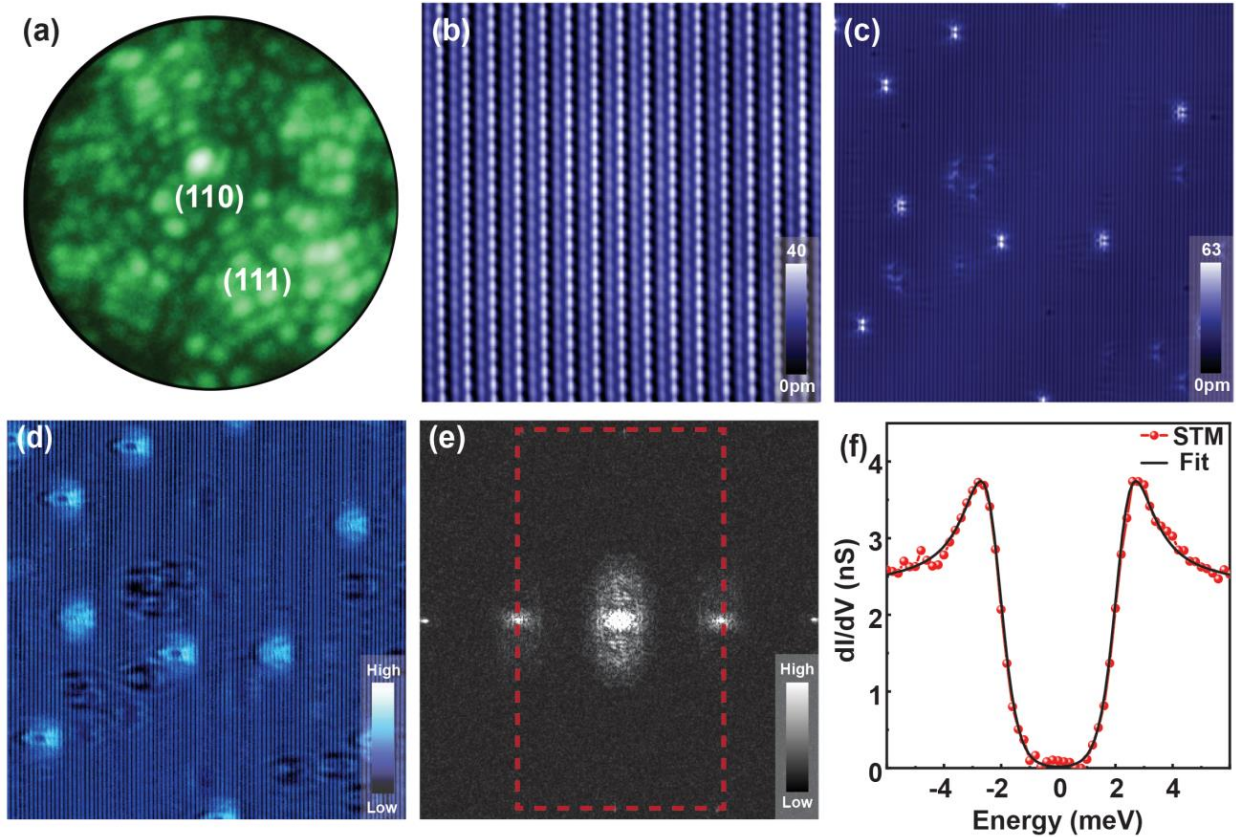

**Figure 5**. Performance of STM. (a) The FIM image of a clean tungsten tip taken with the tip – screen distance of 15 mm and the tip voltage of 7.7 kV. (b) High resolution topographic image of type-II Weyl semimetal WTe$_2$ taken at T ~ 4.2 K with V = +5mV and I =100pA. The scan size is 10 × 10 nm$^2$. (c) Topographic image of WTe$_2$ measured at T ~ 4.2 K with V = 200 mV and I = 100 pA in a large scan size of 60 × 60 nm$^2$. (d) Differential conductance map, dI/dV($r$, E = -110 meV) taken in the same field of view as in (c) revealed the quasiparticle scattering interference pattern around impurities in real space. (e) dI/dV($q$, E = -110 meV) from Fourier analysis of the differential conductance map in (e). The red dash represented the first Brillouin zone. (f) The superconducting gap of δ-NbN thin film taken at T ~ 1.5K with lock-in modulation = 0.5 meV (red) and the fitted curve by Dyne function (black).

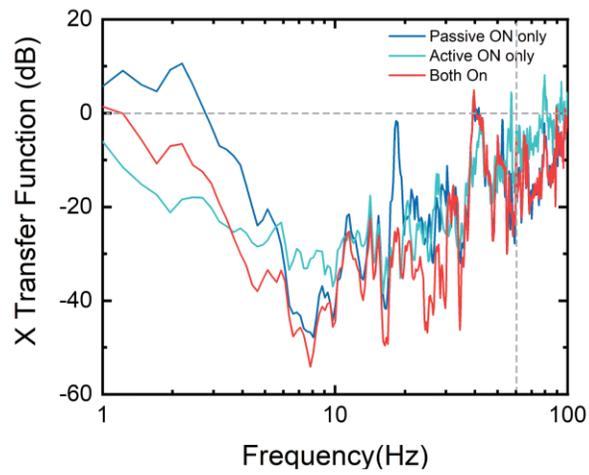
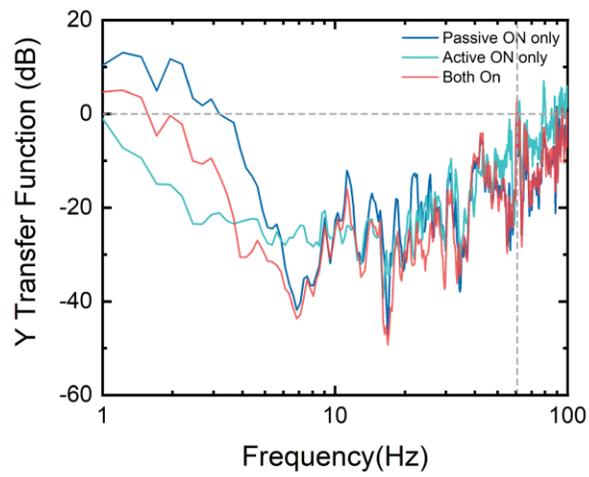
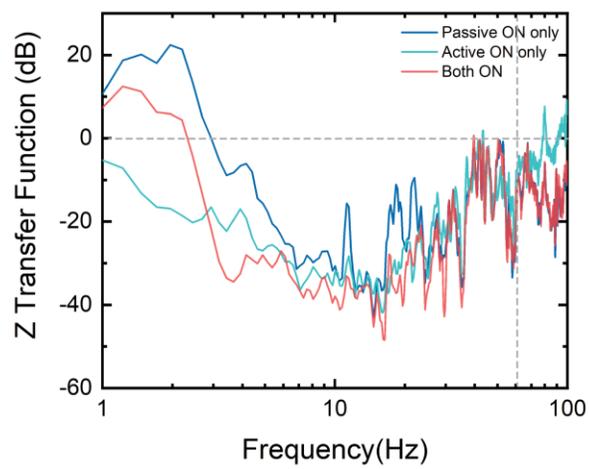

**Supplemental Figure.** Comparison of the transfer functions of our hybrid vibration isolation system along (a) X-direction, (b) Y-direction, and (c) Z-direction, respectively.